\documentstyle[twocolumn,floats,aps,psfig]{revtex}
\begin{document}

\draft
\twocolumn[\hsize\textwidth\columnwidth\hsize\csname
@twocolumnfalse\endcsname

\title{Finite size scaling of the bayesian perceptron}
\author{Arnaud Buhot, Juan-Manuel Torres Moreno \and Mirta B. Gordon\cite{cnrs}}
\address{D\'epartement de Recherche Fondamentale sur la Mati\`ere
Condens\'ee,\\ CEA/Grenoble, 17 rue des Martyrs, 38054 Grenoble Cedex 9,
France}
\date{march 20, 1997}

\maketitle

\begin{center}
\begin{abstract}
\parbox{14cm}{
We study numerically the properties of the bayesian perceptron through a
gradient descent on the optimal cost function. The theoretical distribution
of stabilities is deduced. It predicts that the optimal generalizer lies
close to the boundary of the space of (error-free) solutions. The numerical
simulations are in good agreement with the theoretical distribution. The
extrapolation of the generalization error to infinite input space size
agrees with the theoretical results. Finite size corrections are negative
and exhibit two different scaling regimes, depending on the training set
size. The variance of the generalization error vanishes for $N \rightarrow
\infty$ confirming the property of self-averaging.}
\end{abstract}
\end{center}

]

\section{Introduction}

A neural network is able to infere an unknown rule from examples. We
address specifically classification tasks in which a single neuron -a
{\it perceptron}- connected to $N$ input units through weights ${\bf
w} =(w_1, \ldots, w_N)$ attributes labels $\pm1$ given by $\sigma =
\rm sign({\bf w} \cdot {\bbox \xi})$ to input patterns ${\bbox
\xi}=(\xi_1, \ldots, \xi_N)$. A perceptron is able to correctly classify
{\it linearly separable} (LS) problems; the hyperplane orthogonal to ${\bf w}$
separates, in the input space, the patterns
given positive outputs from those given negative outputs.
Given a set $L_{\alpha}$ of $P=\alpha N$ examples, {\it i.e.}
training patterns ${\bbox \xi}^{\mu}$ ($\mu=1,\ldots,P$) with their
corresponding class $\tau^{\mu}$, the process of finding the weights
${\bf w}$ is called {\it learning}. Generally, if the problem is
LS, there is a finite volume of error-free
solutions in weights space. This volume is called {\it version space}.

Most of the learning algorithms proposed so far may be stated as the
minimization of a cost function or empirical risk $E({\bf w}; L_{\alpha})$
in the weights space. The structure of the problem of learning from examples
allows for a statistical mechanics analysis, in which the cost
function is considered as an energy. The performance of the learning
algorithm is calculated through thermal averages with Boltzman distribution in
weights space and quenched averages over all the possible training
sets. In the thermodynamic limit $N \rightarrow \infty$, $P
\rightarrow \infty$ with $\alpha = P/N$ constant, the zero
temperature limit of these averages accounts for the {\it typical}
behaviour of the algorithm. The fraction of training errors
$\epsilon_t$, the generalization error $\epsilon_g$, and the
distribution of distances of the training patterns to the separating
hyperplane $\rho (\gamma)$ can be determined with the assumption of
self-averaging.

The minimization of the number of training errors, called the Gibbs
algorithm, is not the best learning strategy in the case
of LS problems, because it picks up one point in
version space at random. Its typical generalization error (see
(\ref{eq:epsg}) below for the definition) vanishes with the size of
the training set like $\epsilon_g \approx 0.625/\alpha$~\cite{SST}. A
more elaborate strategy is to look for those weights in version space
that maximize the distance of the separating hyperplane to its
closest training patterns~\cite{MSP,Bottou_Vapnik}. These patterns are
called the support vectors~\cite{Vapnik}, and define the
{\it maximal stability} (or maximum margin) perceptron (MSP), that lies at the center of the version space, and whose
generalization error vanishes in the large $\alpha$ regime like
$\epsilon_g \approx 0.5005/\alpha$~\cite{GG1,BSV1}. As the training set only
contains a small fraction of the information
needed to find the underlying rule generating the examples, there
is a lower bound to $\epsilon_g$, given by Bayes decision
theory~\cite{DH1}. Bayesian performance may be implemented by what is
called a commitee machine~\cite{OH1}: through the vote of a large
number of perceptrons trained with Gibbs algorithm. The bayesian generalization error vanishes like $\epsilon_g \approx 0.442/\alpha$, in the limit of large $\alpha$. However, the
convergence of the commitee machine to the optimum is guaranteed only in the limit of an
infinite number of preceptrons. In order to circumvent the complexity
of the commitee machine, several learning algorithms for single
perceptrons, based on the minimization of ad-hoc cost functions, have been
recently proposed~\cite{GG1,BSV1,MF1}. In these approaches, the cost
function is  sought within a given class of functions and has a free
parameter which has to be optimized for each value of $\alpha$, the fraction
of training patterns. The generalization performance of these algorithms is
very {\it close} to the bayesian optimal value. Some of them end up with a
finite fraction of training errors, suggesting that the optimal solution might
lie outside the version space, but it has been established that this is
not the case~\cite{Watkin}. More recently, the cost function that minimizes
the generalization error, was determined through a variational approach, and
it was showed that its minimum endows the perceptron with the optimal,
bayesian, generalization performance~\cite{KC1}.

In this paper, after a somewhat different derivation of the optimal cost
function, we determine the typical distribution of distances of the training
patterns to the bayesian hyperplane, and we present simulation results that
confirm the theoretical predictions.  We find that the optimal bayesian
student lies close to the {\it boundary} of the version space. The finite
size corrections to the generalization error are negative and present two
different scaling behaviours as a function of $\alpha$.

\section{Theoretical results}

In this section, we present an alternative derivation of the optimal
potential for learning linearly separable tasks for completeness and we
deduce the training patterns distance distribution. The theoretical problem
is formulated as follows: the probability that the
classifier assigns class $\sigma$ to pattern ${\bbox \xi}$ after
training with a set $L_{\alpha}$ of $\alpha N$ examples is $P(\sigma
|\{ {\bbox \xi}, {\bf w}, L_{\alpha}\}) = \Theta(\sigma {\bf w}
\cdot {\bbox \xi}) P({\bf w}|L_{\alpha}) P({\bbox \xi})$, where
$\Theta(x)$ is the Heaviside function. In general the {\it posterior}
probability $P({\bf w}|L_{\alpha})$ is determined through the
minimization of a cost function $E({\bf w}; L_{\alpha})$ which
depends on the training set. In order to derive the properties of a training
algorithm minimizing a cost function, it is useful to introduce a fictitious
temperature $1/\beta$, and consider the finite temperature probability

\begin{equation}
\label{eq:P(w/L;beta)}
P({\bf w}|L_{\alpha};\beta) = p({\bf w}) \frac{e^{-\beta E({\bf
w}; L_{\alpha})}}{Z(L_{\alpha};\beta)},
\end{equation}

\noindent where $p({\bf w})$, called the {\it prior} probability
density, allows to impose constraints to the weights, and
$Z(L_{\alpha};\beta)$ is the partition function

\begin{equation}
\label{eq:partition}
Z(L_{\alpha};\beta)=\int \exp [-\beta E({\bf w}; L_{\alpha}) ]
p({\bf w}) d{\bf w}.
\end{equation}

\noindent The {\it typical} behaviour of any intensive quantity
$X({\bf w})$ is obtained under the assumption of self-averaging through the
quenched average over {\it all}
the possible training sets $L_{\alpha}$ of the same size $\alpha$, in
the thermodynamic limit $N \rightarrow \infty$ (taken at constant
$\alpha = P/N$) and in the zero temperature limit:

\begin{eqnarray}
\label{eq:average}
\ll X \gg & = & \lim_{\scriptstyle N\rightarrow \infty} \int P(L_{\alpha}) dL_{\alpha} \\
\nonumber & & \left[ \lim_{\scriptstyle \beta \rightarrow \infty} \int X({\bf w})
P({\bf w}|L_{\alpha};\beta) d{\bf w} \right].
\end{eqnarray}
\noindent where $\ll \dots \gg$ stands for the double average, over the
weights ${\bf w}$ and the training sets $L_{\alpha}$.

If the cost function has a unique minimum ${\bf w}^*(L_{\alpha})$ (this
may not be the case, as happens when the cost function is the number
of training errors), then $P({\bf w}|L_{\alpha}) = \delta({\bf w}-{\bf
w}^*(L_{\alpha}))$. In this case, the average between brackets in
(\ref{eq:average}) is reduced to $X(L_{\alpha},N)$, which is a random
variable that depends on the particular training set realization through
${\bf w}^*(L_{\alpha})$. The width of its probability distribution function
is expected to vanish in the thermodynamic limit, {\it i.e.} all the
training sets endow the perceptron with the same properties, with
probability one. This property is called self-averaging.  As a consequence,
$X(L_{\alpha},N)$ may be calculated by averaging over all the possible
training sets to get rid of the particular training set realization. The
replica method of statistical mechanics has been developped to cope with the
averages over so called quenched variables which in this case correspond to
the realizations $L_{\alpha}$.

Consider the paradigm of learning a LS rule from
examples: for each pattern ${\bbox \xi^{\mu}}$, a {\it teacher}
perceptron of weight vector $\bf v$ defines the corresponding target
$\tau^{\mu}={\rm sign}({\bf v} \cdot {\bbox \xi^{\mu}})$. As usual,
we assume that the $P=\alpha N$ training patterns are independently
selected with a probability density function $P({\bbox \xi^{\mu}})$, and
that the cost function the {\it student's} weights $\bf w$ have to
minimize is an additive function of the examples,

\begin{equation}
\label{eq:energy}
E({\bf w}; L_{\alpha})  =  \sum_{\mu = 1}^{P} V(\gamma^{\mu}),
\end{equation}

\noindent where the potential $V$ depends on the training pattern
$\mu$ and its class through the stability

\begin{equation}
\label{eq:stability}
\gamma^{\mu}  = \tau^{\mu} \frac{{\bf w} \cdot \bbox{\xi}^{\mu}}{{\sqrt{{\bf
w} \cdot{\bf w}}}}.
\end{equation}

\noindent As the outputs $\sigma^{\mu}$ and $\tau^{\mu}$ are invariant under the
transformations ${\bf w}, {\bf v} \rightarrow a{\bf w}, a'{\bf v}$
with $a, a'>0$, the teacher's and student's weights spaces may be
restricted to the hyperspheres ${\bf w}^2 = N$ and ${\bf v}^2 = N$
respectively without any loss of generality. Most training algorithms can be
cast in the form (\ref{eq:energy}). If the minimum of (\ref{eq:energy}) is
unique and $V(\gamma)$ is differentiable, the weights ${\bf w}$ can be
obtained by a gradient descent. This is not the case for Gibb's algorithm,
whose potential is the non-differentiable error-counting function
$V^G(\gamma)=\Theta(-\gamma)$.

The generalization error $\epsilon_g(\bf{w})$ is the probability that
a pattern, chosen at random with the same probability density as the
training patterns, be misclassified by the student perceptron.
Its typical value depends on the overlap $R=\ll{\bf v}\cdot{\bf
w}/N\gg$ between the student and the teacher weight vectors $\bf{w}$
and $\bf{v}$,

\begin{equation}
\label{eq:epsg}
\epsilon_g \equiv \ll\epsilon_g({\bf w})\gg=\frac{1}{\pi}\arccos R.
\end{equation}

We assume that the training patterns are identically distributed
random variables whose components have zero mean $<\xi_i^{\mu}>=0$ and unit
variance
$<\xi_i^{\mu}\xi_j^{\nu}>=\delta_{\mu \nu} \delta_{ij}$. The free
energy per neuron is averaged over the training sets with the replica method
under the assumption of {\it replica symmetry}, which will be shown to be
stable. The extremum conditions on the free energy that determine the
overlap $R$ are~\cite{GG1}:

\begin{mathletters}
\label{eqs:R1_R2}
\begin{eqnarray}
\label{eq:R1}
1-R^2  =  2 \, \alpha \int_{-\infty}^{\infty}
H\left(\frac{-Rt}{\sqrt{1-R^2}}\right)\left(\lambda(t;c)-t\right)^2 Dt \, ,\\
\label{eq:R2}
R  =  2 \, \alpha
\int_{-\infty}^{\infty}
\exp\left(-\frac{t^2}{2(1-R^2)}\right) \frac{\left(\lambda(t;c)-t\right)dt}
{2\pi \sqrt{1-R^2}} \, ;
\end{eqnarray}
\end{mathletters}

\noindent with $Du = \exp (-u^2/2 ) du/\sqrt{2\pi}$ and $H(t) =
\int_{t}^{\infty} Du = (1/2) {\rm erfc} (t/\sqrt{2})$. The parameter
$c$ is the $\beta \rightarrow \infty$ limit of $\beta (1 - q)$, where
$q$ is the overlap between two solutions in the student's space. If the cost
function (\ref{eq:energy}) has a single global minimum, $q \rightarrow 1$
and $c$ is finite. The function $\lambda(t;c)$, determined by the saddle
point equation of the free energy for $\beta \rightarrow \infty$, minimizes
$W(\lambda) = V(\lambda) + (\lambda-t)^2/2c$ with respect to
$\lambda$. For cost functions having continous derivatives $\lambda(t;c)$
satisfies:

\begin{equation}
\label{eq:lambda}
t = \lambda + c \  \frac{dV}{d\lambda}(\lambda).
\end{equation}

\noindent The solution to (\ref{eqs:R1_R2}) has to verify the necessary
condition for local stability of the replica symmetric solution~\cite{BSV1}:

\begin{equation}
\label{eq:replica_stab}
2 \, \alpha \int_{-\infty}^{+\infty} Dt
H\left(\frac{-Rt}{\sqrt{1-R^2}}\right) (\lambda'(t;c)-1)^2 < 1,
\end{equation}

\noindent where $\lambda' = \partial \lambda/\partial t$. It has recently
been shown that (\ref{eq:replica_stab}) can only be satisfied if
(\ref{eq:lambda}) is invertible, which imposes~\cite{Bouten}:

\begin{equation}
\label{eq:V_second}
c \, V'' \equiv c \, \frac{d^2 V}{d \lambda^2}>-1,
\end{equation}

If $V(\lambda)$ is known,  $\epsilon_g$ can be calculated through
the solution of equations (\ref{eqs:R1_R2}). Instead of solving this
direct problem, we are interested in finding the {\it best}
potential within the class of functions having continuous
derivatives. Instead of using Schwartz inequality as in~\cite{KC1}, we show
that a staightforward functional minimization of $R$ leads to the same
result. As only the product $\beta V$ appears in the partition function
(\ref{eq:partition}), we can multiply the potential $V$ and the
temperature $1/\beta$ by the same constant $a>0$ leaving
$Z(L_{\alpha};\beta)$ invariant. This transformation changes $c
\rightarrow c/a$ in (\ref{eq:lambda}) and (\ref{eq:V_second}),
leaving $R$ unchanged. Thus, we may impose $c=1$ throughout without any loss
of generality, which amounts to choosing the energy units.

A further simplification arises from considering $R$ as a functional
of $V$ through $g(t) \equiv \lambda(t) - t$. For then we can write:

\begin{equation}
\label{eq:g(t)}
g(t)  =  - \frac{dV}{d\lambda}\left(\lambda(t)\right),
\end{equation}

\noindent where $\lambda(t)$ is the solution to (\ref{eq:lambda}).
Equations (\ref{eqs:R1_R2}) and (\ref{eq:replica_stab}) become respectively:

\begin{mathletters}
\label{eqs:f_h}
\begin{eqnarray}
\label{eq:f}
1- R^2 & = & \alpha f(R,g), \\
\nonumber & \equiv & 2 \, \alpha \int_{-\infty}^{\infty} g^2(t)
H \left(\frac{-Rt}{\sqrt{1-R^{2}}}\right) Dt \\
\label{eq:h}
R & = & \alpha h(R,g)\\
\nonumber  & \equiv & \frac{\alpha}{\pi} \int_{-\infty}^{\infty}g(t)\exp
\left(-\frac{t^2}{2(1-R^2)}\right) \frac{dt}{\sqrt{1-R^2}}.
\end{eqnarray}
\end{mathletters}

\begin{equation}
\label{eq:rep_stabbis}
2 \, \alpha \int_{-\infty}^{\infty}
\left(g'(t)\right)^2 H \left( - \frac{Rt}{\sqrt{1-R^2}}\right)Dt<1.
\end{equation}

\noindent Given $\alpha$, equations (\ref{eqs:f_h}) and
(\ref{eq:rep_stabbis}) must be simultaneously verified by the
function $g(t)$ that maximizes $R$. We look for
solution $g(t)$ that minimizes (\ref{eq:h}), with (\ref{eq:f}) considered as a
constraint, introduced through a Lagrange multiplier $\eta$. As it is not
easy to impose inequality (\ref{eq:rep_stabbis}) as supplementary constraint, we
minimize $R = \alpha h(R,g) + \eta [ 1 - R^2 - \alpha  f(R,g) ]$ and we will
show that our result is consistent, {\it i.e.} that the $g(t)$ obtained does
indeed verify conditions (\ref{eq:V_second}) and (\ref{eq:rep_stabbis}). The
function $g(t)$ that maximizes $R$ satisfies $\delta R/\delta g(t)=0$,
which implies $\delta h/\delta g = \eta \delta f/\delta g$, where
$\delta(...) / \delta g$ stands for the functional derivative of $(...)$
with respect to $g(t)$. It is straightforward to deduce the expression for
$g$:

\begin{equation}
\label{eq:g}
g(t) = \frac{\eta^{-1}}{2 \sqrt{2\pi (1-R^2)}}
\frac{\exp \left( -\frac{R^2 t^2}{2(1-R^2)}\right)}
{H\left(-\frac{Rt}{\sqrt{1-R^{2}}}\right)},
\end{equation}

\noindent where $\eta$ and $R$ depend implicitly on $\alpha$. After
introduction of (\ref{eq:g}) into (\ref{eqs:f_h}), we find the
solutions $R(\alpha) \equiv {\cal R}$ and $\eta(\alpha)$:

\begin{mathletters}
\label{eqs:R_eta}
\begin{equation}
\label{eq:R}
\frac{{\cal R}^2}{\sqrt{1-{\cal R}^2}} =  \frac{\alpha}{\pi}
\int_{-\infty}^{\infty} Dt \frac{\exp
\left(-\frac{t^2 {\cal R}^2}{2}\right)}{H(-{\cal R}t)},
\end{equation}
\begin{equation}
\label{eq:eta}
\eta^{-1}(\alpha) = 2 \ \frac{ 1 -{\cal R}^2}{{\cal R}}.
\end{equation}
\end{mathletters}

\noindent They determine, through (\ref{eq:g}), the function $g$ for
each value of $\alpha$:

\begin{equation}
\label{eq:g_of_alpha}
g(t; \alpha) = T^2 \frac{d}{dt} \ln H\left(-\frac{t}{T}\right),
\end{equation}

\noindent where we wrote $g(t;\alpha)$ to stress the $\alpha$
dependence, and $T^2 \equiv (1 -{\cal R}^2)/{\cal R}^2$. It is
straightforward to verify that $g(t;\alpha)$ satisfies
the stability condition (\ref{eq:rep_stabbis}) for all $\alpha$,
justifying our assumption of replica symmetry. A comparaison of
(\ref{eq:R}) with previous results~\cite{OH1,GT1} shows that
${\cal R}(\alpha) = \sqrt{R_G(\alpha)}$, where $R_G$ corresponds to
Gibb's algorithm. The same equation relates the Bayesian generalizer
to Gibb's algorithm, as was demonstrated by Opper and
Haussler~\cite{OH1} with a method that makes explicit use of the
commitee machine architecture.
The potential $V(\lambda)$ may be obtained by integration of
(\ref{eq:g(t)}):

\begin{equation}
\label{eq:V(lambda)}
V(\lambda) = \int_{t(\lambda)}^{+\infty} g(t')
\,\left(1+\frac{dg(t')}{dt'}\right) dt',
\end{equation}

\noindent where $t(\lambda)$ is given by the inversion of
$\lambda=t+g(t;\alpha)$, and we imposed that $V(+\infty)=0$. This optimal
potential endows the perceptron with Bayesian generalization performance and
depends {\it implicitly} on the size of the training set through $T$. It
presents a logarithmic divergence $V(\lambda) \approx -T^2\ln(\lambda)$ for
$\lambda \rightarrow 0^+$. As
$V(\lambda)= \infty$ for negative stabilities to ensure that
$\lambda(t)$ is single valued, the optimal weight vector lies {\it
within} the version space. For $\lambda \rightarrow \infty$, $V(\lambda)
\approx T^3\exp(-\lambda^2/2T^2)/\lambda$. Thus, the range of the
potential decreases for increasing values of $\alpha$ and vanishes as
$\alpha \rightarrow \infty$ showing that the most relevant patterns for
learning are located within a narrow window, on both sides of the student's
hyperplane, whose width shrinks like $T$ for increasing $\alpha$ ($T \sim
1/\alpha$ for $\alpha \gg 1$). With this cost function, the optimal
generalizer may be found by a simple gradient descent, with neither the need
to train an infinite number of perceptrons for implementing a commitee
machine, as was suggested by Opper and Haussler~\cite{OH1}, nor to
determine a large number of 'samplers' of the version space, as
proposed by Watkin~\cite{Watkin}.

Once the potential is known, it is straightforward to calculate the
distribution of stabilities of the training set:

\begin{equation}
\label{eq:def_rho}
\rho(\gamma)=\ll\frac{1}{P}\sum_{\mu}\delta(\gamma-\gamma^{\mu})\gg.
\end{equation}

\noindent Its general expression is~\cite{GG1}:

\begin{equation}
\label{eq:stab}
\rho(\gamma) = 2 \int_{-\infty}^{\infty}Dt \,
H\left(-\frac{t}{T}\right) \delta\left[\lambda(t)-\gamma\right]
\end{equation}

\noindent with $\lambda(t) = t + g(t;\alpha)$. In terms of $t(\gamma)$,

\begin{equation}
\rho(\gamma) = \sqrt{\frac{2}{\pi}}
\exp\left(\frac{-t^2(\gamma)}{2}\right)
H\left(\frac{-t(\gamma)}{T}\right)
\frac{dt}{d\gamma}(\gamma)
\end{equation}

\noindent which depends on $\alpha$ through $T$. In the present case, as all
the patterns have positive stabilities, $\rho(\gamma)$ is the distribution
of the distances of the training patterns to the student's hyperplane.
Distributions obtained through a numerical inversion of
$\lambda(t)$, for several values of $\alpha$, are plotted in
fig~\ref{stability}~\cite{K1}. The density of patterns is exponentially small,
$\rho(\gamma) \approx [T/\pi \gamma]
\exp[-T^2/2({\cal R} \gamma)^2]$ at small distance to the hyperplane. It
increases with $\gamma$ up to a maximum at $\gamma_{M}(\alpha)$. At larger
$\gamma$ there is a crossover to a gaussian distribution, $\rho(\gamma)
\approx (\sqrt{2/\pi}) \exp(-\gamma^2/2)$ identical to the teacher's one.
Both $\gamma_{M}(\alpha)$ and the crossover distance  get closer to the
hyperplane
with increasing $\alpha$. In the large $\alpha$ limit,
both quantities vanish like $1/\alpha$, with $\gamma_{M}\approx
1.769/\alpha$. Thus, the region of disagreement between the
student's and the teacher's distributions decreases for
increasing size of the training set. In the limit $\alpha \rightarrow
\infty$, the bayesian distribution is identical to the teacher's one.

It is worthwhile to compare the present results with the MSP, whose weight
vector is the one with maximal distance from all the hyperplanes that
define the version space. The corresponding distribution $\rho(\gamma)$
presents
a gap for $\gamma<\kappa$, and a $\delta$ peak at $\gamma=\kappa$,
which is precisely half the smallest width of the version space. In the
large $\alpha$ limit, $\kappa \approx 1.004/\alpha$ is smaller than
$\gamma_M$. The fact that the bayesian student has patterns at vanishing
distance from the hyperplane, and has most patterns at distances larger than
$\kappa$, allows us to conclude that its weight vector lies close to the
boundary of the version space. It has been shown~\cite{Watkin} that the
bayesian weight vector is the barycenter of the (strictly convex) version
space. Our result means that the barycenter of the version space is far
from its center, which is rather surprising, and might indicate that the
version space is highly non-spherical. Notice that the teacher weight
vector lies even closer to the version space boundary, as it has a finite
distribution of stabilities for all $\gamma > 0$. This explains why some
potentials recently
proposed~\cite{GG1,MF1} may reach a generalization error lower than the MSP,
in spite of the fact that they find a solution {\it outside} the version
space, {\it i.e.} without correctly learning the complete training set.

\section{Simulation results}
\label{sec:simulation}

The theoretical results of the preceding section were obtained in the
thermodynamic limit, $N \rightarrow \infty$, $P \rightarrow \infty$,
with $\alpha=P/N$ finite. In this section we present results of
thorough numerical simulations that confirm very nicely the
theoretical predictions, and are precise enough to determine the finite
size corrections.

We describe first our implementation of the learning procedure. Given
a training set, the optimal student is found by a gradient
descent on the cost function (\ref{eq:energy}) with potential
(\ref{eq:V(lambda)}). In practice, only the derivative of the
potential is needed, and we do not need to perform the integration in
(\ref{eq:V(lambda)}). As $dV/d\lambda$ is the function $-g(t;\alpha)$
defined by equation (\ref{eq:g_of_alpha}), evaluated at
$t=t(\lambda)$ given by (\ref{eq:lambda}), we only have to invert the
equation $\lambda(t) = t+g(t;\alpha)$. We calculated numerically
$dV/d\lambda$ for each value of $\alpha$ considered. As the
optimal potential diverges for negative stabilities, the minimization
has to be started with a weight vector ${\bf w}(0)$ already inside
the version space. In our simulations, we determined ${\bf w}(0)$ by
minimization of the cost function (\ref{eq:energy}) with potential
$V(\lambda) = 1-{\rm tanh}(\beta \gamma /2)$, in which the value of $\beta$
has to be optimally tuned ~\cite{GG1}. We used the implementation
called Minimerror ~\cite{RG1}, that finds the best value of $\beta$
together with the weights ${\bf w}(0)$ through a deterministic
annealing. Starting from ${\bf w}(0)$, the weights are iteratively
modified through

\begin{mathletters}
\label{eq:w(k+1)}
\begin{eqnarray}
\label{eq:w(k+1),a}
{\bf w} & = & {\bf w}(k) - \epsilon(k)\,\delta {\bf w},    \\
\label{eq:w(k+1),b}
\delta {\bf w} & = & \sum_\mu
\frac{dV}{d\lambda}(\gamma^\mu;\,\alpha\,)\,\tau^\mu\,{\bbox \xi}^\mu,   \\
\label{eq:w(k+1),c}
{\bf w}(k+1) & = & N \frac{{\bf w}}{{\bf w} \cdot {\bf w}}
\end{eqnarray}
\end{mathletters}

\noindent where $\gamma^\mu$ is the stability (\ref{eq:stability}) of
pattern $\mu$. Actually, the derivative $\partial E/\partial {\bf w}$
has two terms, and only one of them is taken into account in equation
(\ref{eq:w(k+1),b}). The neglected term, that contributes to keep
${\bf w} \cdot {\bf w}$ constant only to first order in $\epsilon$,
has been replaced by the normalization (\ref{eq:w(k+1),c}). A
straightforward calculation shows that the component of $\delta
{\bf w}$ (eq. (\ref{eq:w(k+1),b})) orthogonal to ${\bf w}(k)$,
$\delta {\bf w}_\perp \equiv \delta {\bf w} - {\bf w}(k) \delta
{\bf w} \cdot {\bf w}(k)/N$, is proportional to $\partial E/\partial
{\bf w}$. Thus, at convergence, $\delta {\bf w}_\perp^2 \equiv
\delta {\bf w}_\perp \cdot \delta {\bf w}_\perp$ vanishes.
Actually, the stopping condition in all our simulations was
$\delta {\bf w}_\perp^2 \leq 10^{-14}$.

The variable learning rate $\epsilon(k)$, introduced to speed-up the
convergence, is determined as follows: at each iteration, we
calculate (\ref{eq:w(k+1),a}) for three different values of
$\epsilon$: $\epsilon(k-1)/2$, $\epsilon(k-1)$, and
$5\,\epsilon(k-1)$. The value $\epsilon(k-1)/2$ should prevent the
oscillations that may appear for too large learning rates, whereas
$5\,\epsilon(k-1)$ allows to accelerate the convergence in regions
where the potential is flat. At each iteration, we keep for $\epsilon(k)$
the value that minimizes $\delta {\bf w}_\perp^2$. With this procedure, the
initialization of $\epsilon$ is irrelevant; we used $\epsilon (0) = 10^{-2}$
in all our tests.

In our simulations, we determined the generalization error
$\epsilon_g(\alpha,N)$ and the distribution of stabilities
$\rho(\gamma;\alpha,N)$ as a function of $N$ and $\alpha$.
Given $\alpha$ and $N$, we generated training sets $L_\alpha$ of
$P=\alpha N$ binary patterns. The components of the input patterns
${\bbox \xi}^\mu$ of each training sample were chosen at random with
probability $p(\xi_i^\mu=1) = p(\xi_i^\mu=-1) = 1/2$ for all $1 \leq
i \leq N$ and $1 \leq \mu \leq P$. The corresponding outputs
$\tau^\mu={\rm sign}({\bf v} \cdot {\bbox \xi}^\mu)$ are determined
by a randomly selected teacher of normalized weights ${\bf v}$ (${\bf
v} \cdot {\bf v} = N$). We made simulations for $\alpha = 1, 2, 4, 6,
8, 10$ and $14$, and we considered at least seven different values of
$N$ for each value of $\alpha$. Each training set was learnt with the
optimal potential using (\ref{eq:w(k+1)}), as explained before. The
overlap between the obtained normalized weights ${\bf w}^*(L_{\alpha})$ and the
teacher's weights ${\bf v}$, $R(L_{\alpha},N) = {\bf w}^* \cdot {\bf v}/N$,
determines the generalization error of the student perceptron,
$\epsilon_g(L_{\alpha},N) = \arccos[R(L_{\alpha},N)]/\pi$.

We determined the generalization error for each pair $(\alpha,N)$, averaged
over $M(\alpha,N)$ training sets, $\epsilon_g(\alpha,N) =
\sum_{\{L_{\alpha}\}} \epsilon_g(L_{\alpha},N) /M(\alpha,N)$. The number of
samples $M(\alpha,N)$ was chosen large enough to have a good precision in
the extrapolation to $1/N \rightarrow 0$. Values of $M(\alpha,N)$ ranging
from $500$ to $20\,000$ (the larger number of samples corresponding to the
smaller values of $P = \alpha N$) were used. Most of the simulations were
done on a parallel computer that allows for $64$ samples to be processed
simultaneously. The obtained values of
$\epsilon_g(\alpha,N)$ are represented on figure \ref{fig:eg(1/N)} as
a function of $1/N$. All the investigated values of $\alpha$ show the
same behaviour, and only some of them are reported on the figure for
reasons of clarity. The generalization errors are linear in $1/N$
because for each $\alpha$ we only considered values of $N$ large
enough that the second order corrections in $1/N$ be negligible. The
fits to the numerical results extrapolate correctly to the
theoretical values $\epsilon_g({\alpha})$ obtained in the thermodynamic
limit $N \rightarrow \infty$, $P \rightarrow \infty$ with $\alpha =
P/N$ constant. The finite size corrections are {\it negative},
meaning that in finite dimension the expected generalization error is {\it
lower} than predicted by the theory. This result can
be understood if one considers the information content of the
training set instead of its size. As the number of possible training
patterns is $2^N$, the training set carries (on the average) a
fraction of information $\alpha N/2^N$ which, at constant $\alpha$,
is larger the smaller $N$. Moreover, given $\alpha$, there
is always a value $N_{\alpha}$ large enough that $\alpha N_{\alpha} > 2^{N_{\alpha}}$, {\it i.e.} such that {\it all} the possible patterns belong to the training set. One expects that $\epsilon_g(\alpha,N_{\alpha})=0$, and that
$\epsilon_g(\alpha,N)$ increases smoothly for increasing $N$ to reach
$\epsilon_g(\alpha,\infty)$ from below.

The  variance of the generalization error, $\sigma^2_g(\alpha,N) =
\sum_{\{L_{\alpha}\}} (\epsilon_g(L_{\alpha},N)-
\epsilon_g(\alpha,N))^2 /M(\alpha,N)$, is represented as a function
of $1/N$ on figure \ref{fig:var(1/N)} for all the values of $\alpha$
considered. The fact that all the lines extrapolate to zero shows
that, in the thermodynamic limit, the distribution of
$\epsilon_g(L_{\alpha},N)$ is a delta function: any randomly selected
training set corresponding to the same $\alpha$ endows the perceptron with
the same typical
generalization error, with probability one. In other
words, the hypothesis of self-averaging,
underlying the statistical mechanics calculations, is correct.

At finite size, the average generalization error and its variance
depend on $P = \alpha N$. To first order in $1/P$, we may write:

\begin{eqnarray}
\epsilon_g(\alpha,N) & = & \epsilon_g(\alpha,\infty) -
\phi(\alpha)/P, \\
\sigma^2_g(\alpha,N)  & = & \psi(\alpha)/P.
\end{eqnarray}

\noindent The behaviour of $\phi(\alpha)$ and $\psi(\alpha)$,
displayed on figures \ref{fig:slopdeg} and
\ref{fig:slopvar}, presents a crossover at $\alpha \simeq 2$, {\it
i.e.} in the neighbourhood of the perceptron's capacity. At large
$\alpha$, $\phi(\alpha)$ is constant and $\psi(\alpha)$ decreases
smoothly, whereas at small $\alpha$, both quantities increase with
$\alpha$. Thus, as a function of $P$, finite size corrections to
$\epsilon_g$ vanish slower at $\alpha \lesssim
2$ than at large $\alpha$. This is the reason why we needed a larger number
of samples for low $\alpha$ in our simulations.

As $N$ decreases, the mean value of the generalization error distribution,
$\epsilon_g(\alpha,N)$, shifts towards lower values, proportionally to
$1/N$. However, the broadening of the distribution, $\sigma_g(\alpha,N) \sim
1/\sqrt{N}$, overcompensating this effect. Thus, in spite of the negative
correction to $\epsilon_g(\alpha,\infty)$ at finite $N$, there is a finite
probability that a particular trained perceptron generalize worse than the
theoretical prediction.

The distributions of stabilities follow the same trends as the
generalization error. Histograms, determined with some of our
results, are compared to the theoretical density distributions, on
figures \ref{fig:staba4} and \ref{fig:staba6}. On figure
\ref{fig:staba4}, numerical results for both the student and the
teacher perceptrons, are displayed. Although not clearly visible on
the figure, the finite size teacher has less patterns at small
distances to the separating hyperplane, the tail of the distribution
being slightly higher, than the theoretical distribution. These
discrepancies are much smaller than the finite size effects on the
student perceptrons, which exhibit an increase of the pattern density
closer to the hyperplane, with a corresponding depletion of the peak
at $\gamma_M$. These efects are enhanced at smaller $N$, as may be
seen on figure \ref{fig:staba6}.

\section{Conclusion}

In this paper, we presented numerical simulations of the simplest
neural network, the perceptron, learning optimally a linear
separation task from examples. They confirm the theoretical
predictions and present interesting finite size scaling behaviours.

After a derivation of the optimal learning potential, we deduced the
theoretical distribution of distances of the learned patterns to the
separating hyperplane, $\rho(\gamma)$. Surprisingly, the optimal
student is predicted to be close to the boundary of the version space
instead of being near of its center, as currently believed.

We presented extensive numerical simulations with the aim of clarifying to
which extent the theoretical results, which predict the typical behaviour of
the generalization error and the distribution of stabilities in the
thermodynamic limit, are valid for finite size systems. In particular, the
numerically determined distribution of stabilities shows that finite size
optimal
perceptrons lie even closer to the version space boundary than the
theoretical prediction for $N \rightarrow \infty$. The extrapolation of the
generalization error $\epsilon_g$ to $1/N \rightarrow 0$ averaged over a
large number of samples, confirm the theoretical predictions with very high
accuracy. The variance of $\epsilon_g$ vanishes in that limit, showing that
all the training sets endow the perceptron with the same generalization
error, with
probability one. This is just what is meant by the hypothesis of
self-averaging underlying the replica approach, which is thus
numerically validated.

At finite $N$ the mean generalization error is {\it smaller} than the
theoretical value. As the argument that allows to understand such result
is independent of any learning scheme, for it takes into account only
the information content of the training set, we expect it to be also
valid for statistical mechanics predictions of $\epsilon_g$
for other learning algorithms. However, it is worth to point out that the
width of the generalization error distribution grows with decreasing $N$
faster than the shift of the mean value.

As a function of $\alpha$, $\epsilon_g(\alpha,N)$ shows two different
scaling regimes, depending on whether $\alpha>2$ or $\alpha<2$.  The
crossover at $\alpha_c=2$ might be correlated to the perceptron capacity. As
below $\alpha_c$ any training set is expected to be linearly separable, it
seems likely that the
generalization error presents a different scaling at $\alpha < \alpha_c$.
Theoretical calculations of finite size corrections remain to be done, to
clarify the observed scaling regimes.

Although the simulations were done for binary random input vectors, the
behaviour of the generalization error should be the same for continuous
input vectors whose components have zero mean and unit variance, as the
theoretical results only depend on the two first moments of the pattern
distribution. It would be interesting to see whether the observed
cross-over at $\alpha \approx 2$ persists in this case.

\newpage

\begin{figure}
\caption{ \label{stability} Distribution of distances of the training
patterns to the bayesian separating hyperplane for different values of
$\alpha$.}

\vskip 0.5cm

\caption{ \label{fig:eg(1/N)}
Average generalization error vs. $1/N$. Error bars are not visible at the
scale of the figure. Lines are least squared fits to the numerical data,
which are extrapolated to $1/N = 0$. Full symbols correspond to the
theoretical values.}

\vskip 0.5cm

\caption{ \label{fig:var(1/N)} Variance of the generalization error vs.
$1/N$. Lines are least squared fits to the numerical data.}
\vskip 0.5cm

\caption{ \label{fig:slopdeg} Slope of the finite size corrections to the
generalization error.}
\vskip 0.5cm

\caption{ \label{fig:slopvar} Slope of the finite size variance of the
generalization error.}

\vskip 0.5cm

\caption{ \label{fig:staba4} Theoretical and numerical ($N = 100$)
distribution of stabilities for the optimal student and the teacher
for $\alpha = 4$.} 
\vskip 0.5cm

\caption{ \label{fig:staba6} Theoretical and numerical ($N = 20$ and $65$)
distribution of stabilities for the optimal student for $\alpha = 6$.}
\end{figure}

. \\


\begin{thebibliography}{99}
\bibitem{K1} After completion of this work, we learned that O. Kinuchi
derived a similar expression for the distribution of stabilities
(Unpublished).
\end{thebibliography}
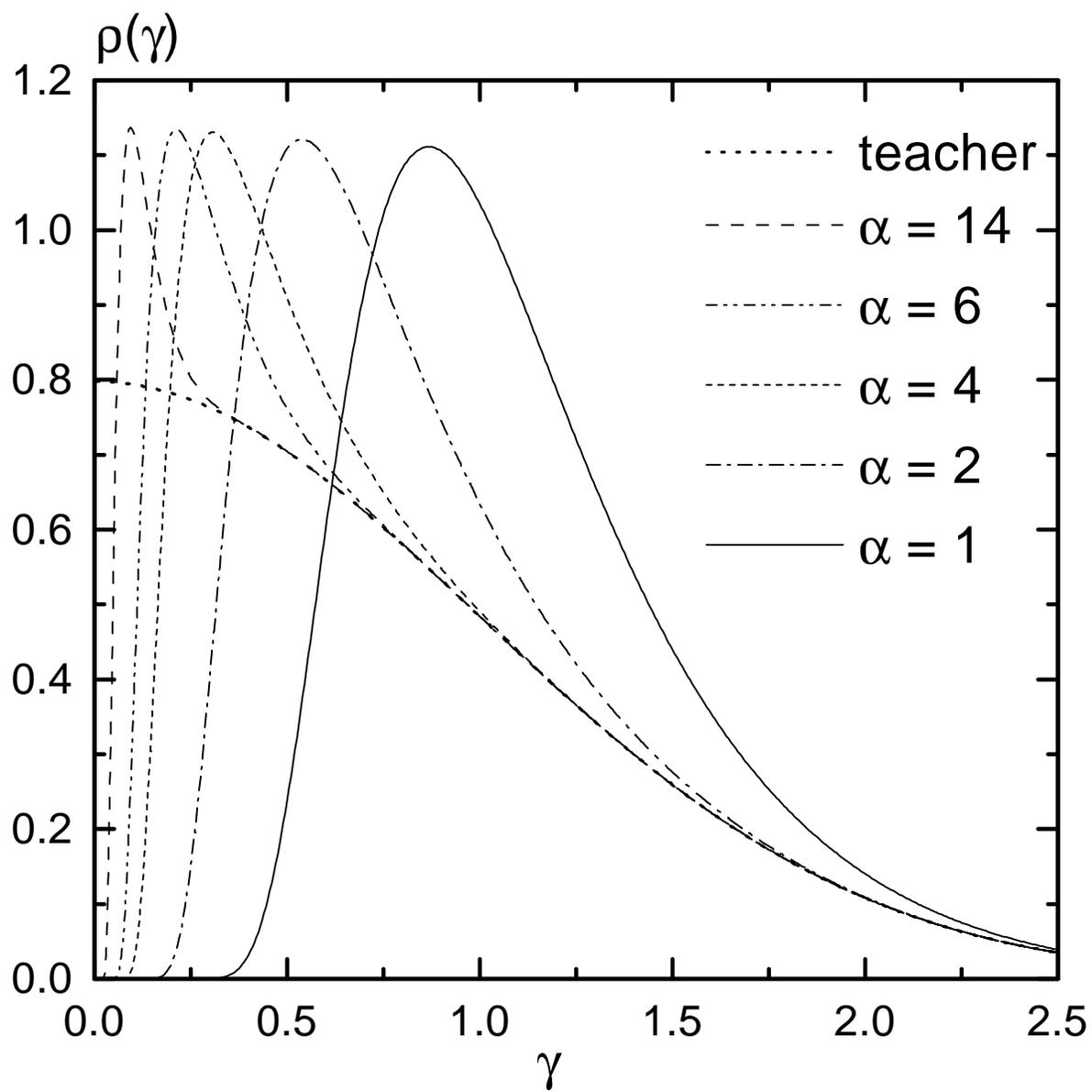
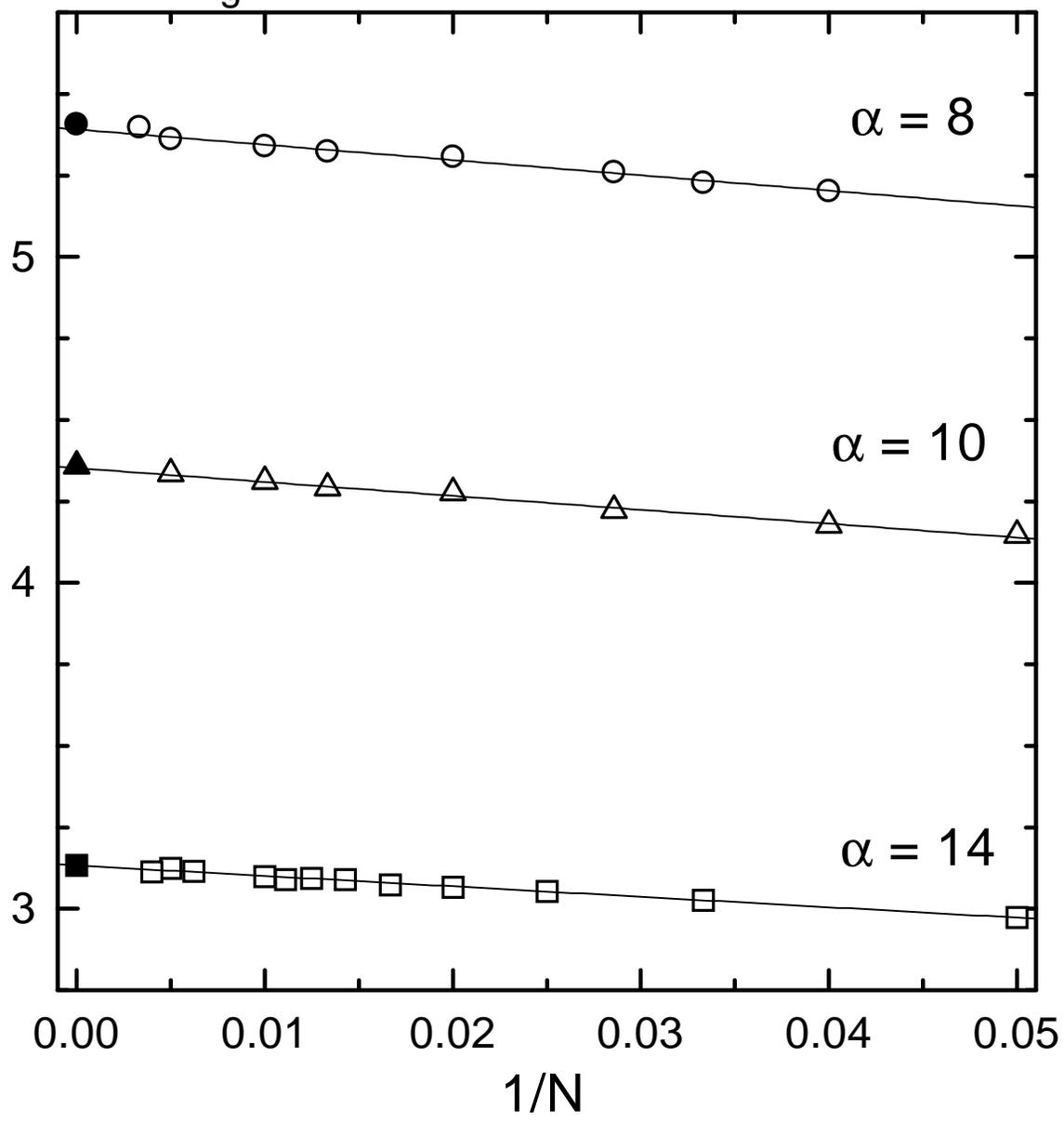
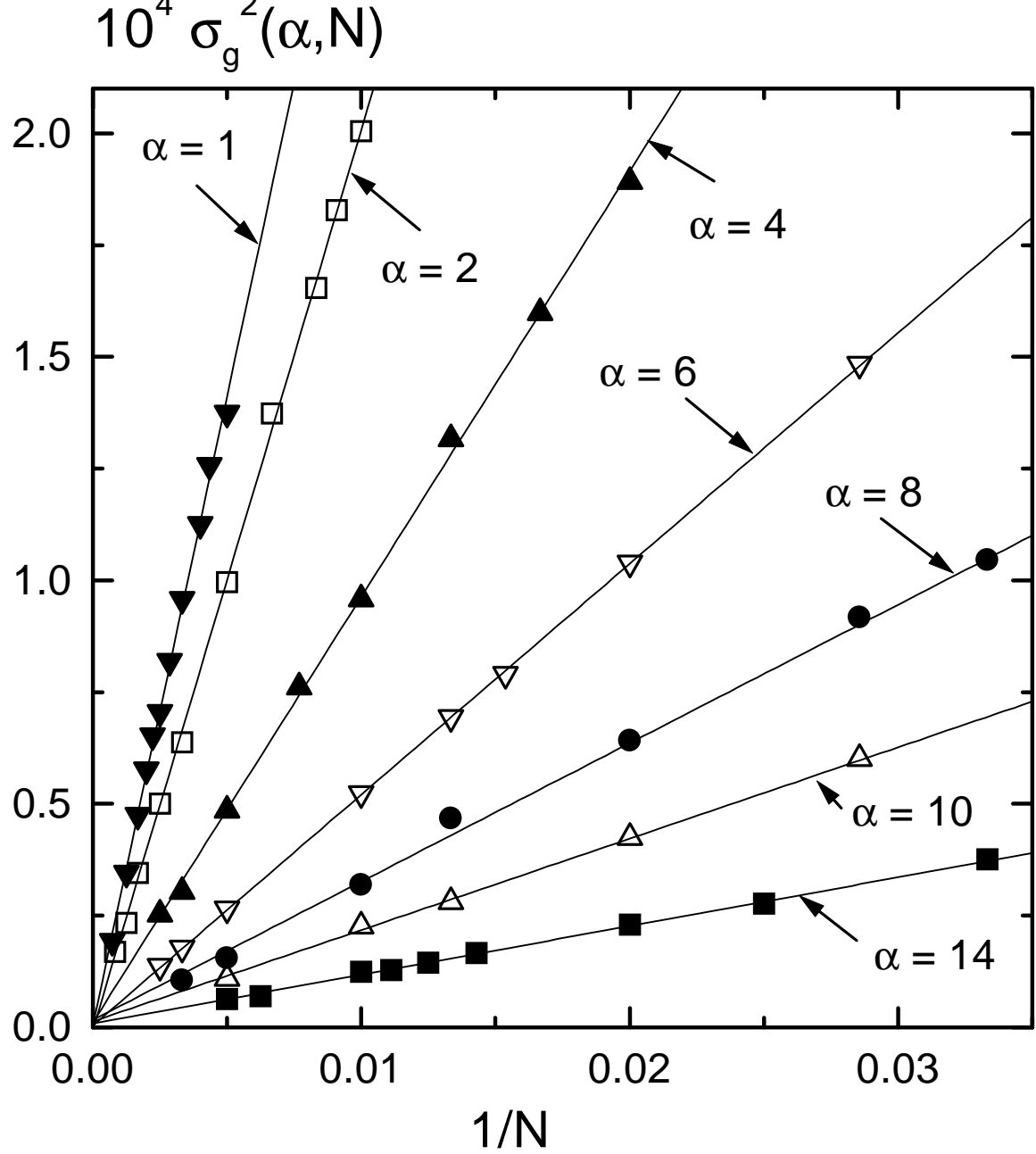
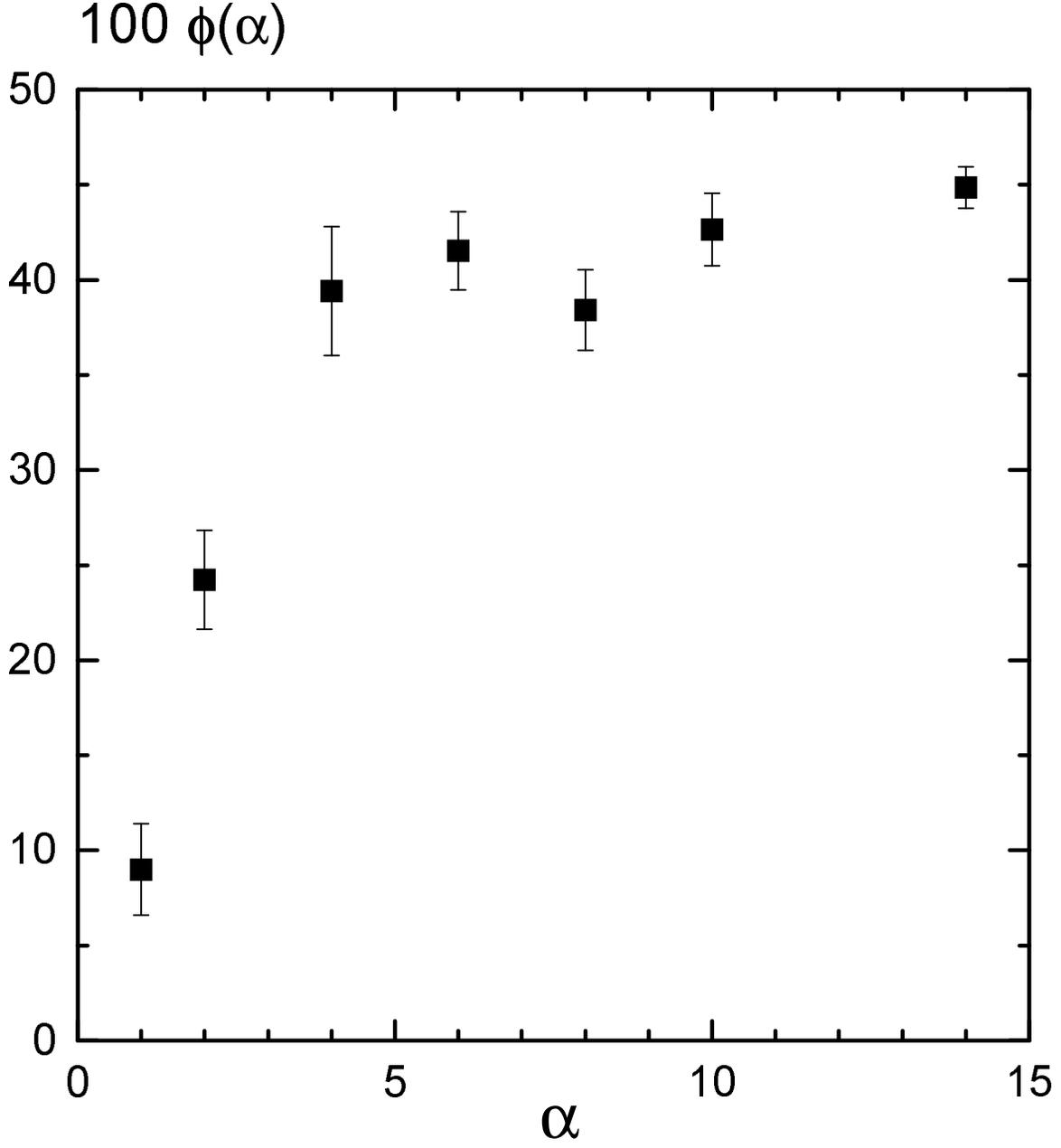
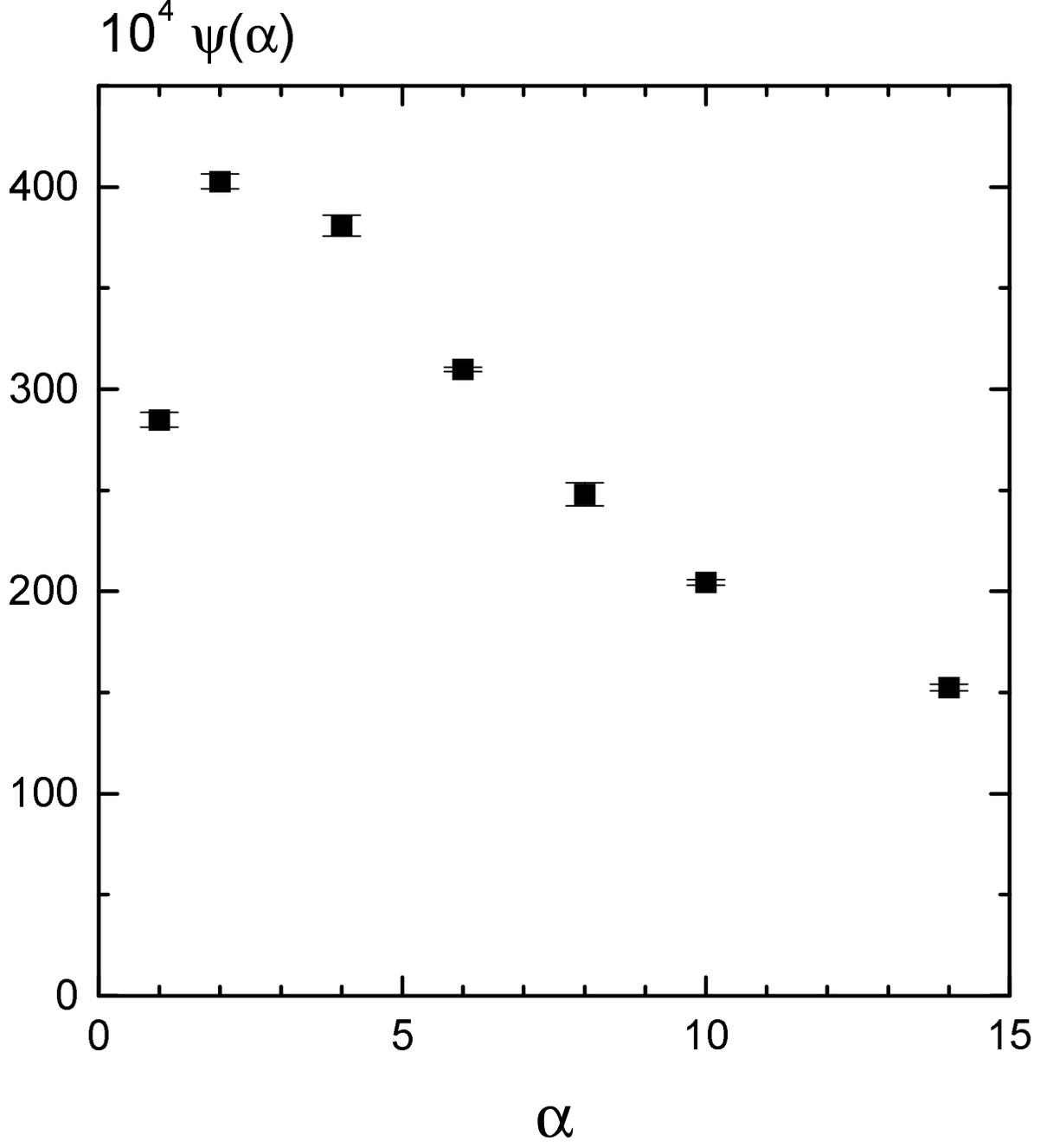
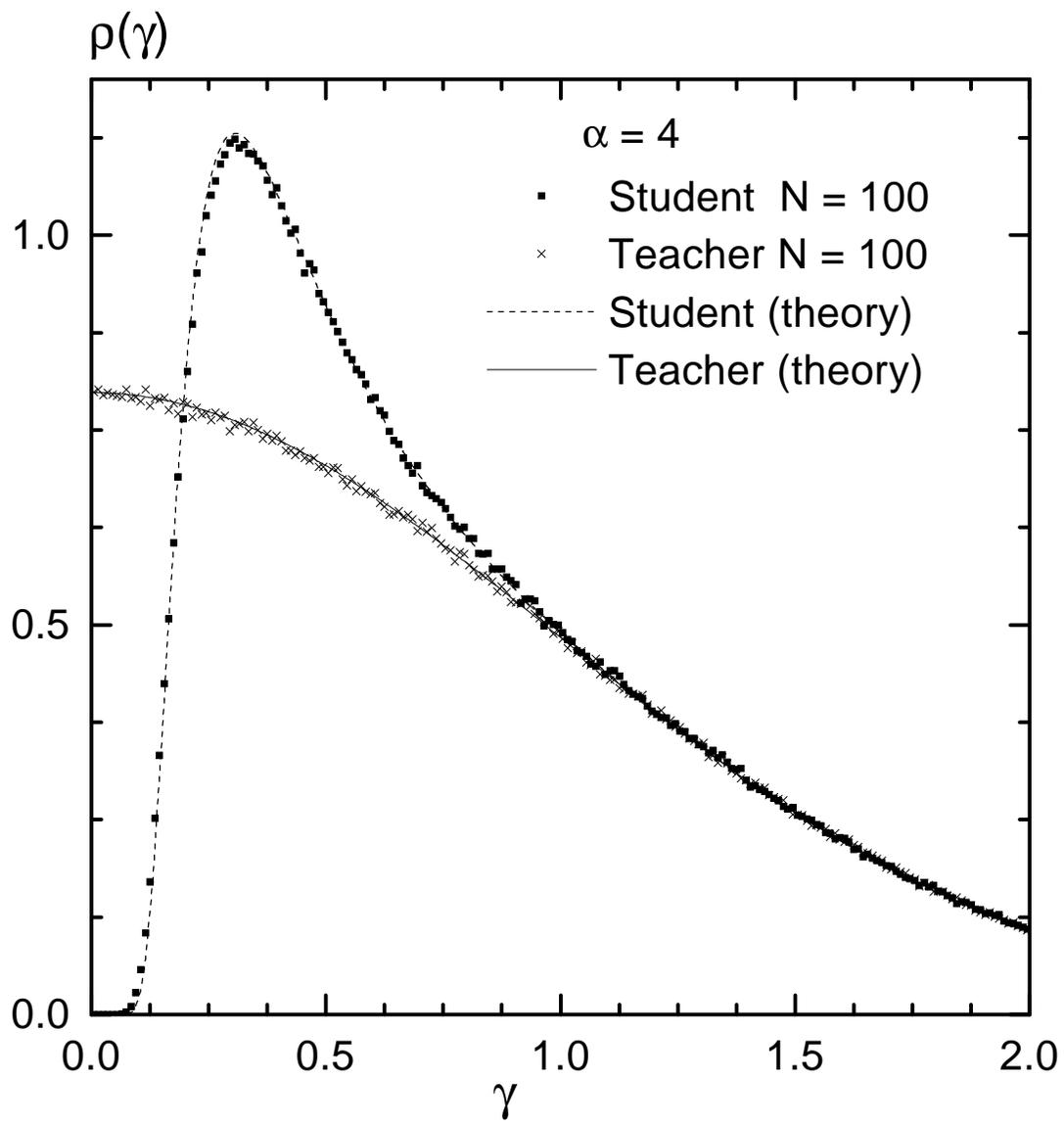
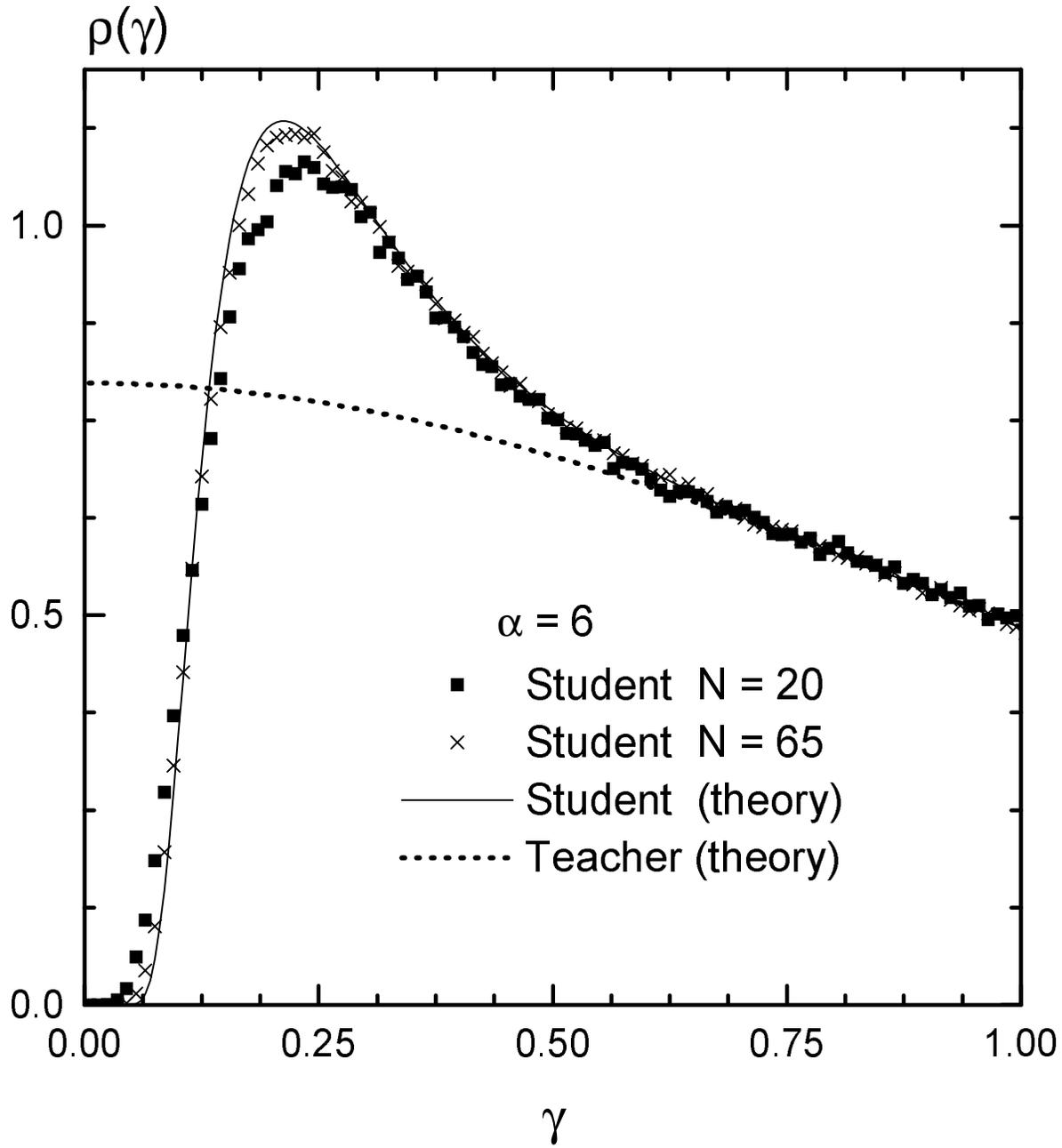
\end{document}